# Probing Dynamics at Interfaces: Options for Neutron and X-ray Spectroscopy

Maikel C. Rheinstädter[1]*, Tilo Seydel[1], Bela Farago[1] and Tim Salditt[2]

[1]Institut Laue-Langevin, 6 rue Jules Horowitz, BP 156, 38042 Grenoble Cedex, France
[2]Institut für Röntgenphysik, Georg-August Universität, Geistrasse 11, 37037 Göttingen, Germany

**Abstract**
Inelastic neutron and X-ray scattering experiments on surfaces and interfaces are a challenging topic in modern physics. Particular interest arises regarding surfaces and interfaces of soft matter and biological systems. We review both neutron and x-ray spectroscopic techniques with view at their applicability to these samples. We discuss the different methods, namely neutron triple-axis, backscattering and spin-echo spectroscopy as well as x-ray photon correlation spectroscopy, in the context of planar lipid membrane models as an example. By the combination of the different methods, a large range in momentum and energy transfer is accessible.

**Introduction**
An outstanding problem of modern condensed matter physics relates to the question how structure, thermodynamics, phase transitions and molecular motions change from the bulk values when the spatial dimensions are reduced. In recent years, a growing interest has arisen in studying dynamics at surfaces and interfaces in as large a range of time scales as possible. While most spectroscopic techniques, as e.g. nuclear magnetic resonance or dielectric spectroscopy, are limited to the center of the Brillouin zone at Q=0 and probe the macroscopic response, neutron and x-ray scattering experiments give the unique access to microscopic dynamics at length scales of e.g. intermolecular distances. To enlarge the Q-$\omega$ range to a maximum, several experimental techniques from neutron and x-ray scattering have to be combined. Figure 1 gives an overview of the accessible Q-omega range of different spectroscopic techniques. By combining neutron triple-axis or time-of-flight, backscattering and spin-echo spectrometers, an energy range from about 50 meV (thermal triple-axis or time-of-flight) down to sub-$\mu$eV (spin-echo), corresponding to timescales from about 0.1 ps to 100 ns, is accessible. X-ray photon correlation spectroscopy (XPCS) even extends this range down in the neV range and beyond (detectable motions slower than about 50 ns). Neutron scattering gives access to length scales ranging from intermolecular distances of about 3 Å up to several hundred Å. Topics of interest are for instance the glass transition at the surface, the test of theoretical predictions derived from continuum mechanics, polymer surface dynamics or the dynamics of biological model systems such as planar lipid membranes. To solve these issues, several experimental challenges have to be met: The minuteness of the inelastic signals necessitates a sample preparation and experimental set-ups specially adapted for inelastic experiments. In this paper we focus on the application of different inelastic scattering techniques to the study of lipid membranes as prominent examples of low dimensional systems. The understanding of dynamics in these model membranes is of fundamental interest in biophysics. We mainly discuss energy resolved neutron techniques, but also include a short section on XPCS.

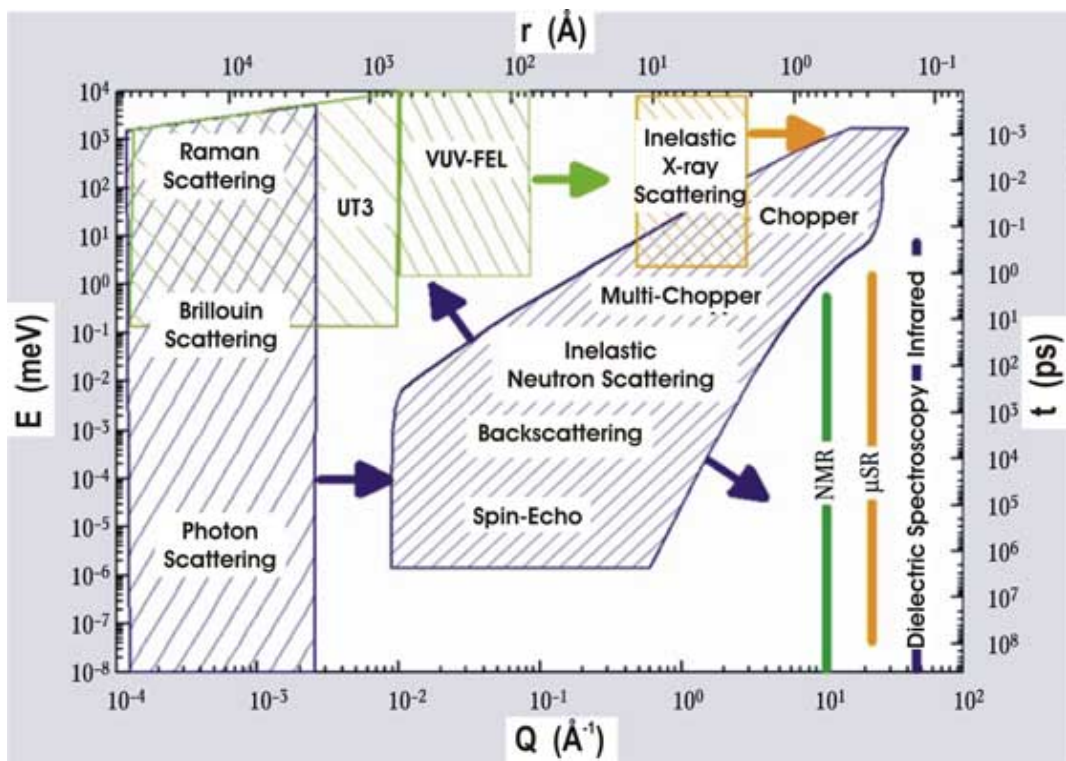

**Figure 1:** Accessible Q-ω range for different spectroscopic techniques including inelastic neutron and X-ray scattering (adapted from ESS).

The paper is organized as follows: In the next section we discuss basic properties and preparation techniques of lipid membranes for inelastic neutron and x-ray experiments. The experiments have been done using neutron triple-axis, backscattering and spin-echo spectrometers as well as x-ray photon correlation spectroscopy. Each technique will be described in a dedicated section, with a focus on the experimental details.

**Lipid Membranes**
Phospholipid membranes are intensively studied as simple model systems to understand fundamental structural and physical aspects of their much more complex biological counterparts (1). Neutron scattering can contribute to the elucidation of the molecular structure, as is well documented in the literature, as well as to the understanding of molecular and supramolecular dynamics of lipid bilayers. Dynamical properties are often less well understood in biomolecular systems, but are important for many fundamental biomaterial properties, e.g. elasticity properties and interaction forces. Lipid membrane dynamics on small molecular length scales for instance determines or strongly affects functional aspects, like diffusion and parallel or perpendicular transport through a bilayer. Here we discuss inelastic neutron scattering for studies of the collective motions of the acyl-chains on different length and time scales. Molecular vibrations, conformational dynamics and "one particle" diffusion in the plane of the bilayer can be studied by a number of different spectroscopic techniques covering a range of different time scales such as incoherent inelastic neutron scattering or nuclear magnetic resonance or dielectric spectroscopy. Contrarily, few experimental techniques, namely coherent inelastic neutron scattering or inelastic x-ray scattering, are able to elucidate the short range collective motions mentioned above.

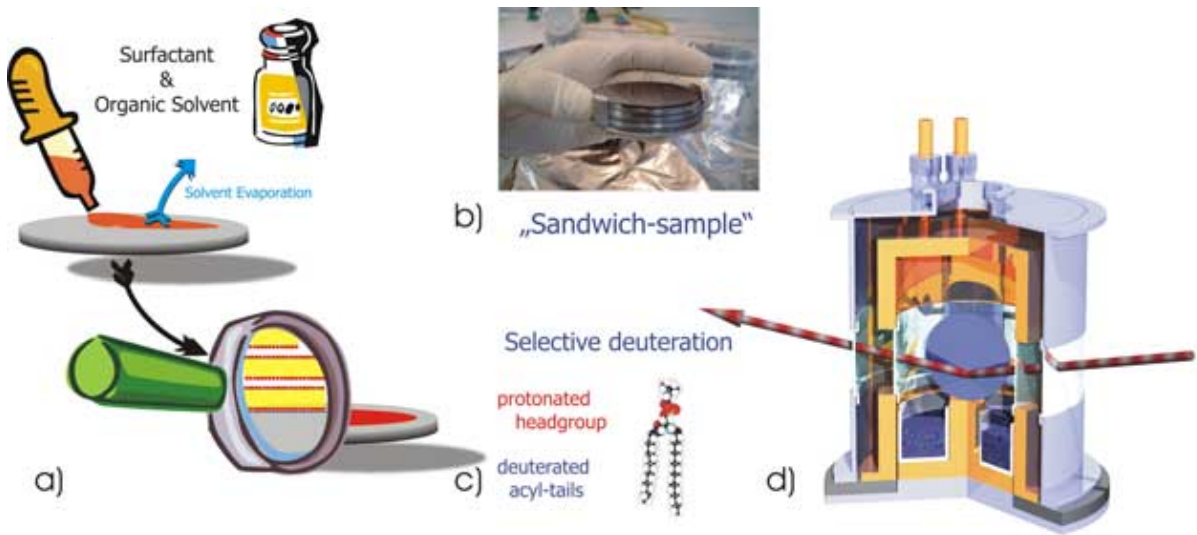

**Figure 2:** (a) Sketch of the sample preparation. (b) Photograph of the "sandwich sample" used for the neutron experiments. (c) By selective deuteration, the collective motions of the acyl tails are enhanced over other contributions to the inelastic scattering section. (d) Schematic of the humidity can that allows controlling temperature and humidity of the bilayers.

Because of the minuteness of the inelastically scattered signals, the preparation of appropriate samples and experimental set-ups is challenging. For the neutron triple-axis, backscattering and spin-echo experiments with beam sizes of several centimeters and almost negligible absorption, we have prepared solid supported membrane stacks on 2'' Silicon wafers. For XPCS measurements which are carried out using micrometer sized pinholes as coherence filters, much smaller samples can be used (5x10 $mm^2$ in the present case). As a high purity model systems, we have chosen the zwitterionic phospholipids dimyristoyl-phosphatidylcholine (DMPC, Avanti Polar Lipids,AL). The solid supported, highly oriented membrane stacks were prepared by spreading a solution of typically 35 mg/ml lipid in trifluoroethylene/chloroform (1:1) on 2'' silicon wafers (or correspondingly smaller amounts if smaller substrates were used), followed by subsequent drying in vacuum and hydration from $D_2O$ vapor (2). Figure 2 (a) shows a sketch of the sample preparation. Up to twenty such wafers separated by small air gaps were combined for the neutron measurements and aligned with respect to each other to create a "sandwich sample" consisting of several thousands of highly oriented lipid bilayers (total mosaicity of about 0.6 deg), with a total mass of up to 500 mg of deuterated DMPC, see Figure 2 (b) for a photograph of the sample. The samples were kept in a closed temperature and humidity controlled Aluminum chamber during the measurements (Figure 2 (d)). Hydration of the lipid membranes was achieved by separately adjusting two heating baths, connected to the sample chamber and to a heavy water reservoir, hydrating the sample from the vapor phase. Temperature and humidity sensors were installed close to the sample. By selective deuteration of the acyl-chains (Figure 2 (c)), the respective collective motions are strongly enhanced over other contributions to the inelastic scattering cross section. The fully hydrated, deuterated DMPC bilayers undergo the phase transition ("main transition") from the more ordered gel phase into the fluid phase at 21.5 °C. Of particular interest are measurements in the physiological relevant fluid phase.

**Triple-axis spectroscopy**
The concept of triple-axis spectrometry has been very successful in the investigation of collective excitations in condensed matter physics, i.e. phonons and magnons in crystals but has so far not been applied to lipid membranes. Advantages of triple-axis spectrometers are their relatively simple design and operation and the efficient use of the incoming neutron flux to the examination of particular points in $Q,\omega$ space. Figure 3 (a) shows a schematic of a triple-axis spectrometer. By varying the three axes of the instrument, the axes of rotation of the monochromator, the sample and the analyzer, the wave vectors $k_i$ and $k_f$ and the energies $E_i$ and $E_f$ of the incident and the scattered beam, respectively, can be determined. **Q**, the momentum transfer to the sample, and the energy transfer, $\omega$, are then defined by the laws of momentum and energy conservation to **Q**=$k_f$- $k_i$ and $\omega$=$E_i$-$E_f$. The accessible (Q,$\omega$) range of IN12 for a fixed energy of the scattered beam $E_f$ of 10 meV is shown in Figure 3 (b) and covers well the Q-$\omega$ range of the excitations, as we will see later on. It is just limited by the range of incident neutron energies offered by the neutron guide as well as by mechanical restrictions of the spectrometer. The instrumental energy resolution in this configuration is $\Delta$=500 $\mu$eV. By choosing smaller incident energies and energy transfers the energy resolution can be enhanced. The measurements were carried out on the cold triple-axis spectrometer IN12 and the thermal spectrometer IN3 at the high flux reactor of the ILL in Grenoble, France. The combination of cold and thermal neutrons makes accessible an excitation spectrum from about 0.5 up to 30 meV. The use of highly oriented membrane stacks allowed to perfectly aligning the scattering vector **Q** with respect to the lipid bilayers. **Q** can be placed in the plane of the membranes to measure static in-plane correlations, $S(Q_r)$, or in plane dynamics, $S(Q_r,\omega)$. By rotating the sample by 90 deg, **Q** can be set perpendicular to the bilayers to probe interlamellar correlations to determine e.g. the interlamellar spacing and the thickness of the water layer. Triple-axis spectrometry thus offers the possibility to measure reflectivity, $S(Q_r)$ and in-plane dynamics, $S(Q_r,\omega)$, on the same instrument in the same run without changing setup (3)(4). This is an invaluable advantage as the thermodynamic state of the lipid bilayer not only depends on temperature and relative humidity, but also on cooling and heating rates, preparation and thermal history.

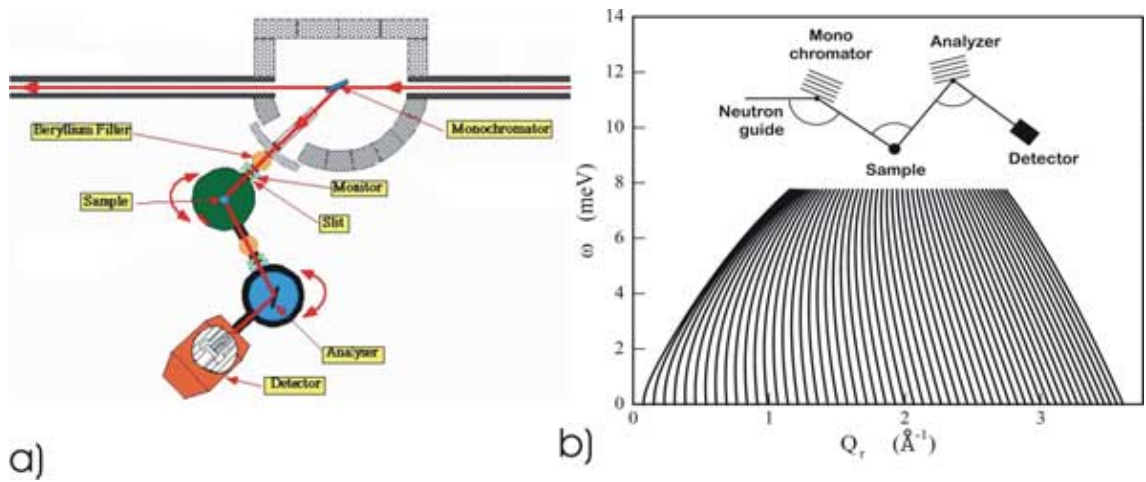

**Figure 3:** (a) Schematic of a triple-axis spectrometer. (b) The accessible Q-$\omega$ range for a typical configuration (fixed $E_f$=10 meV) is hatched.

A typical energy scan is shown in Figure 4 (b). The data were collected at T=20 °C, in the gel phase of the bilayer at Q=1.0 Å$^{-1}$. The inset shows the excitations of the bilayer in the gel and the fluid phase in magnification. Position and width can easily be determined from these well pronounced peaks. Figure 4 (c) shows the dispersion relation in the gel and the fluid phase as measured by several constant Q-scans. The dispersion relation resembles typical dispersions found in fluids.

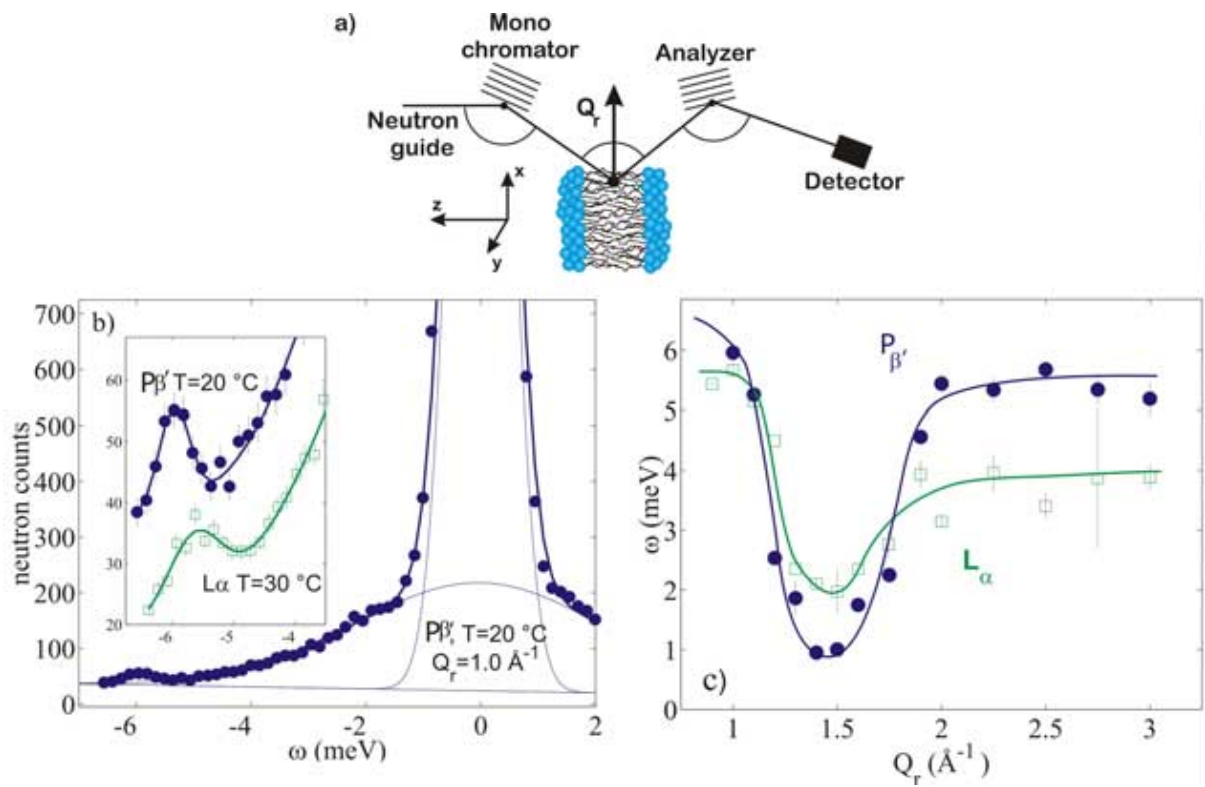

**Figure 4:** (a) Orientation of the bilayers with respect to the spectrometer. (b) Energy scan in the gel phase of DMPC (T=20 °C), measured at $Q_r$=1.0 Å$^{-1}$. The scattering is composed of the central elastic peak, the broad quasi-elastic background, and symmetric satellites. The inset shows a zoom of the satellite peaks in both gel and fluid phase at T=30 °C. (c) Dispersion relations in the gel and the fluid phase of the DMPC bilayer as measured by several constant Q scans. (From (3))

The particular shape of the dispersion relation can qualitatively be explained. The basic scenario is the following: At small $Q_r$, longitudinal sound waves in the plane of the bilayer are probed and give rise to a linear increase of $\omega \sim Q_r$, saturating at some maximum value ("maxon"), before a pronounced minimum $\omega_0$ ("roton") is observed at $Q_0 \cong 1.4$ Å$^{-1}$, the first maximum in the static structure factor $S(Q_r)$ (the inter-chain correlation peak). Qualitatively, this can be understood if $Q_0$ is interpreted as the quasi-Brillouin zone of a two-dimensional liquid (the lipid molecules arrange on a two-dimensional hexagonal lattice). Collective modes with a wavelength of the average nearest neighbor distance $2\pi/Q_0$ are energetically favorable and lead to the minimum. In perfectly ordered crystals, the energy of the acoustic phononic branches goes down to zero at the zone centers. The static and dynamic disorder in the lipid bilayers finally leads to a minimum at finite energy values. At $Q_r$ values well above the minimum, the dispersion relation is dominated by single particle behavior. The dispersion relation can be extracted from Molecular Dynamics (MD) simulations by temporal and spatial Fourier transforming the molecular real space coordinates (5) and shows excellent agreement. While the "maxon" and the high-Q range are energetically higher in the gel than in the fluid phase due to stiffer coupling between the lipid chains in all-trans configuration, $\Omega_0$, the energy value in the dispersion minimum, is actually smaller in the gel phase, roughly analogous to soft modes in crystals. The range at low Q-values is difficult to access by inelastic neutron scattering because of the kinetic restriction, i.e. the dispersion relation of the neutron itself. This restriction does not hold for inelastic x-ray scattering; the "dispersion" of a photon is a straight line with an extremely steep gradient. Inelastic x-ray experiments with meV-resolution are achieved by using a near backscattering geometry, related to the neutron backscattering technique. But because of the backscattering geometry and the very small scattering angles it is very difficult to separate the scattered from the direct beam in the x-ray experiment.

**Neutron Backscattering Spectroscopy**

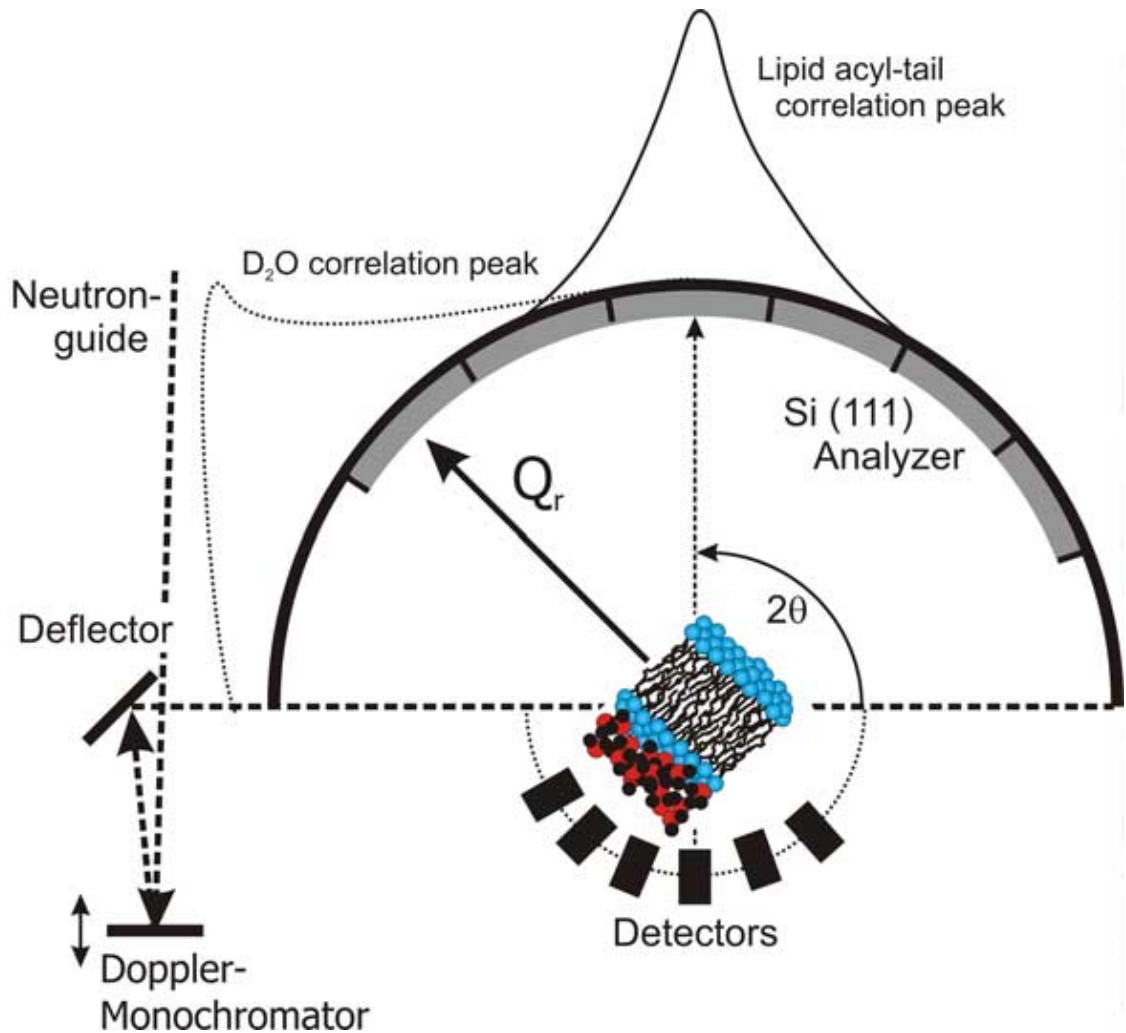

**Figure 5**: Schematic of the backscattering geometry at IN10. Spatially arranged analyzers allow to separately but simultaneously probe the molecular dynamics at different length scales. In our example, the inter-acyl-chain correlation peak in the plane of the membranes is located at 1.4 Å$^{-1}$ (the heavy water correlation peak would occur at 2 Å$^{-1}$).

Backscattering experiments can be technically more easily be achieved using neutrons. In backscattering experiments the scattering function $S(Q,\omega)$ is directly measured analogously to triple-axis spectroscopy, but contrarily to XPCS or neutron spin-echo. The high resolution obtained in exact backscattering is easily shown by computing the first derivative of Bragg's law (6),

$$\frac{\Delta\lambda}{\lambda} = \frac{\Delta d}{d} + \frac{\Delta\Theta}{\tan\Theta}$$

where $\lambda$ is the neutron wavelength, d the monochromator crystal lattice spacing and $\Theta$ the angle of incidence of the neutron beam with respect to the crystal surface. From this equation it becomes clear that the monochromaticity is maximized when the angle of incidence is 90° with respect to the monochromator and analyzer crystal surface of a spectrometer. This geometry can be realized with neutrons by using adequate deflecting disk chopper devices, whereas for x-rays a perfect backscattering setup can only be approximated. For neutrons, exact backscattering has first been realized at the spectrometer IN16 at the ILL (7).

Taking advantage of the relatively large wavelength provided by cold neutrons, the ratio $\Delta\lambda/\lambda$ is particularly favorable, and thus with the backscattering technique a Gaussian energy resolution of 0.9 $\mu$eV FHWM can routinely be achieved, and a resolution of 0.45 $\mu$eV FHWM is possible at a reduced intensity. Neutron backscattering spectrometers typically consist of both a backscattering monochromator and a backscattering analyzer sphere (see Figure 5). The detectors are mounted very close to the sample, and the discrimination between analyzed neutrons and neutrons directly scattered into the detectors is achieved by their time of flight. Therefore, the incident beam is pulsed by a chopper (not shown in Figure 5).

Using the backscattering technique, two basic types of measurements can be performed: With fixed energy-window scans centered at zero energy transfer (FEW-scans), the scattered intensity arising from the sample that is elastic within the instrumental resolution can be recorded as a function of the sample temperature. From FEW-scans, information on the onset and type of molecular mobility in the sample can be inferred. Thus, glass or melting transitions can be clearly identified and assigned to corresponding length scales. The second type of measurement is the energy transfer scan. In backscattering, the energy transfer can be scanned by varying the incident energy. This is done by Doppler-shifting the incident neutron energy through an adequate movement of the monochromator crystal. Hereby, an incident neutron energy shift of about -15$\mu$eV<$\Delta$E<+15$\mu$eV relative to the incident neutron energy can routinely be scanned, with the energy transfer limit only given by the mechanical limit of moving the crystal sufficiently fast. These energy scans correspond to a time range of motion in the sample down to a few nanoseconds. The energy transfer range can be increased to several hundred $\mu$eV by using a heatable monochromator crystal instead of a mechanically moving crystal. This setup is available at IN10 at the ILL.

In view at its entangled geometry, the use of the neutron backscattering technique for probing dynamics at interfaces is challenging. Nevertheless, we have demonstrated the feasibility of backscattering on the lipid membrane sandwich samples. Analogously to the triple axis experiments, we have oriented the samples in the spectrometer to measure at wave vector transfers parallel and perpendicular to the lipid membrane plane, respectively (see Figure 5).

The IN10 analyzers cover an angular range of approximately 20° each, resulting in a rather poor $Q$-resolution, but enhanced sensitivity for even very small inelastic signals. We used six discrete detector tubes of IN10. The broad lipid acyl-chain correlation peak that occurs at $Q_r \cong 1.4$ Å$^{-1}$ was (mainly) detected in one detector tube ('lipid detector'), as depicted in Fig. 1. A $Q$-range of 0.3 Å$^{-1}$<$Q$<1.9 Å$^{-1}$ was simultaneously detected in this set-up to investigate and discriminate molecular dynamics on the different length scales. We performed FEW scans in a temperature range of 100-315 K to map out the transition of the lipids from immobile to mobile as a function of temperature for (a) the scattering vector *Q* placed in the plane of the membranes and (b) perpendicular to the bilayers. While the in-plane component ($Q_r$) in the 'lipid-detector' shows a pronounced freezing transition, there is no distinct T-dependence in the perpendicular direction ($Q_z$). We interpret this in terms of correlated motions, which take place mainly in the plane of the lipid bilayers (in the time and length scales observed). Figure 6(a) shows the in-plane component of the elastic scattering with the measurement in $Q_z$ subtracted as background. We attribute the pronounced freezing step ('immobile' within the resolution window) at 294 K ($Q$ centred at 1.42 Å$^{-1}$) to the main transition of the lipid acyl-chains from the rigid gel phase at low-T into the fluid phase at higher temperatures. When analysing all detectors we find a second transition at about 271 K, mainly in the detector centred at $Q$=1.85 Å$^{-1}$, which tentatively might be attributed to the hydration water of the membrane stacks, i.e. the water layer in between the stacked membranes. Even though the detector is not perfectly centred to the maximum of the static structure factor of water at $Q$=2 Å$^{-1}$ (which is not accessible on IN10), it is positioned to detect a reasonable part of the broad heavy water correlation peak. Within this interpretation, freezing of the hydration water would be lowered by about six degrees as compared to (heavy) bulk water at about 277 K. Fig.3 displays corresponding energy transfer scans. The data have been taken at three different temperatures, for T=250, 290 and 300 K with a typical counting time of about 9 hours per temperature. An elastic peak in the inelastic spectra points to static order at the corresponding length scales, where a fluid system has no order at infinitely long time scales. Even within the very limited statistics, the different dynamics is clearly visible: While the lipid acyl-chains melt between 290 and 300 K, melting at the water position already occurs between 250 and 290 K.

Our experiment gives a first high energy-resolution wave vector-resolved insight into collective lipid membrane dynamics. The dynamical properties of hydration water may be different from those of bulk water because hydrogen bonding to the lipid head groups at the lipid-water interface of the membrane might slow down water rotation and translation. A scenario with gradual freezing of the water molecules, depending on the distance to the water-lipid interface, is also under discussion.

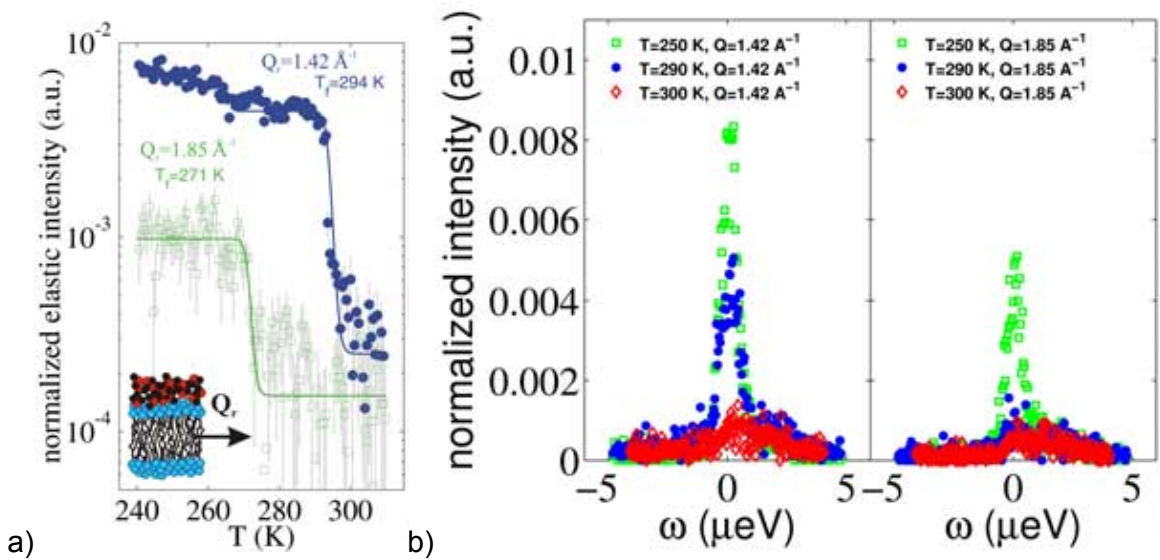

**Figure 6:** (a) In-plane component of the elastic scattering signal. The mobile-immobile transition (within the resolution window) is clearly different for the lipid acyl-chains (T=294 K) and the position of the water correlation peak (T=271 K). Solid lines are guides to the eye. (b) Energy scans at temperatures T=250 K, 290 K and 300 K for the $Q$-values 1.42 Å$^{-1}$ (lipid acyl-chain correlation peak) and $Q$=1.85 Å$^{-1}$. At 290 K, the water signal is already 'mobile' within the experimental energy resolution whereas the lipid acyl-chains are still frozen. (Counting is normalized to monitor)

Further measurements and data analysis are in progress, and with this technique the melting temperatures of the lipid layers and the membrane water will be mapped out. In addition, further experiments will focus on whether the melting is accompanied by a quasi-elastic broadening. This would be an indication of whether or not phase or glass transitions are present in this system.

**Neutron Spin-Echo Spectroscopy**

Neutron Spin Echo (NSE) is a clever way of reaching very high energy resolution with neutrons close to $10^{-5}$ of the incoming neutron energy) without scarifying intensity (14). With the latest development in instrumentation (16)(15) the gap between photon correlation spectroscopy ( X-ray and visible light) is getting smaller and smaller. Offering the possibility to cover the dynamics in systems over 5-7 decades in space (q range) as well as in energy (time space) unavoidably leads us to a more complete description of the physical systems.
With NSE the directly measured quantity is S(q,t)/S(q,0), that is we measure directly the time dependence. Here

$$t \propto \lambda^3 \int B d\mathrm{l}$$

B is the magnetic field of the precessions coils and the integral is taken over the coil length. Usually the magnetic field is varied to cover the time range we are interested in. It is important to note the strong dependence of the time parameter with the wavelength. Essentially the limiting factor for the maximum reachable Fourier time are the residual field inhomogeneities (17), nevertheless choosing longer wavelength still extends the time range. Long wavelength has a second benefit for surface studies. In reflection the angles we are dealing with become bigger for a given q value, thus the sample size to cover the usually large neutron beam (usually in the cm range) is smaller. Presently there exists no dedicated NSE spectrometer for reflectivity studies, the experiment we will describe in the following (13) was performed on the IN15 NSE spectrometer at the ILL.
The sample was a smectic liquid crystal 4-octyl-40-cyanobiphenyl (8CB) with the C8H17 chain deuterated to enhance the neutron contrast.

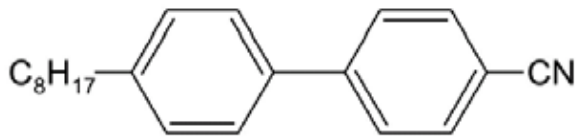

As the detector and the second precession solenoid can be moved only in the horizontal plane, the sample was put in a vertical position. The film was stretched in situ on an aluminum frame, which was as much as possible protected by Cadmium to decrease parasitic background scattering. Due to the vertical position the film thickness was not uniform, going from 0.5 micron on the top to a few microns on the bottom. Fortunately the dynamics we were observing was not (or only very weakly) dependent on the film thickness. The repeat distance of the smectic layers leads to a sharp diffraction peak at q=0.18 Å$^{-1}$. With 9.4 Å incoming wavelength this correspond to 16.8 degree scattering angle. This Bragg peak, with q being perpendicular to the film, corresponds to the projection of the time averaged neutron scattering length density in the film plane. As such is was expected, and experimentally was found too, to give only elastic scattering. The film dynamics is expected to be seen only if q has a component parallel to the film. As the NSE spectrometer uses a not too monochromatic beam ($\Delta\lambda/\lambda = 11\%$ FWHM) but a relatively tight angular collimation (about 0.5 degree), the most convenient way to move off the Bragg peak is to rotate the sample. Here again the long wavelength is of a advantage. With the 16.8 degree scattering angle we could easily rotate the sample +/- 5 degree without obscuring the incoming or outgoing beam. On Figure 7 the scattering geometry is represented with the sample turned horizontal for better visibility (18).

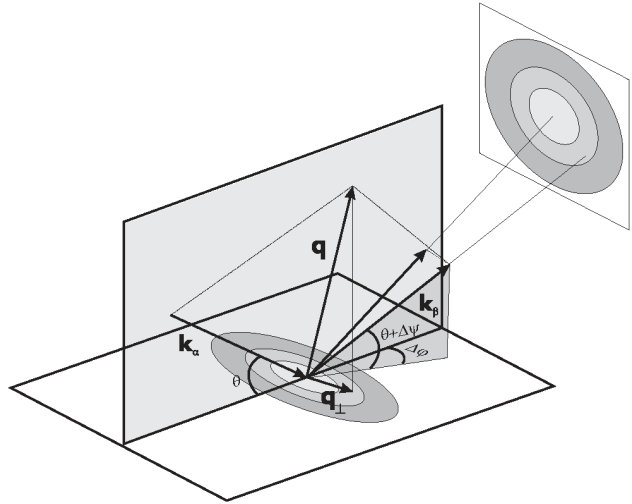

**Figure 7:** Sketch of the spin-echo scattering geometry.

IN15 has a 2D detector, thus the in plane q component can be calculated as follows:

$$q_\perp = \frac{2\pi}{\lambda}\sqrt{\left(\cos(\Delta\varphi)\cos\left(\frac{\vartheta_0}{2}+\Delta\psi-\omega\right)-\cos\left(\frac{\vartheta_0}{2}+\omega\right)\right)^2 + \left(\sin(\Delta\varphi)\right)^2},$$

where $\vartheta_0$ is the scattering angle, $\omega$ is the sample rotation, $\Delta\psi, \Delta\varphi$ are angles which correspond to the pixel positions on the 2D detector relative to the center. The detector pixels were grouped into three groups around constant $q_\perp$ values as shown below.

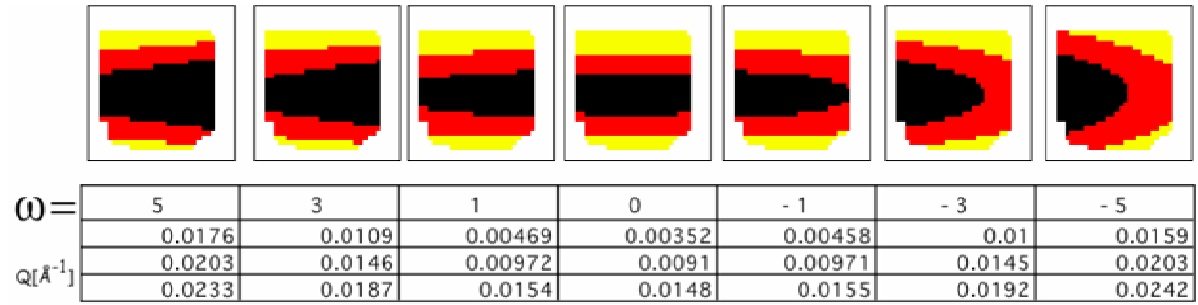

Typical NSE decay curves are shown in Figure 8. The q=0 curve corresponds to the Bragg peak and within accuracy it is elastic. All the other curves could be well described with a stretched exponential with β=0.6 exponent.

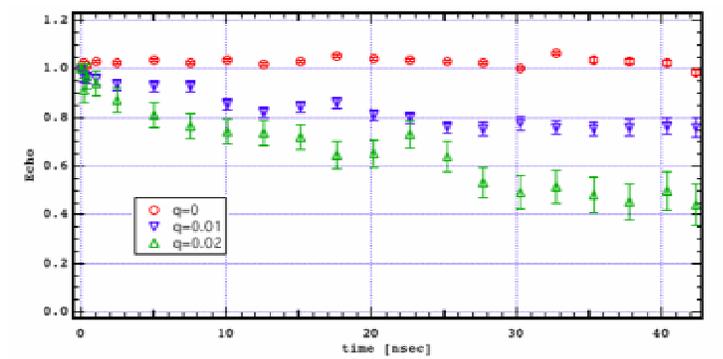

**Figure 8:** NSE decay curves for different q-values.

Taking into account the bending elasticity of the film (13)(18) NSE measurements extend the q range to a new regime of previous XPCS results with good agreement with theory.

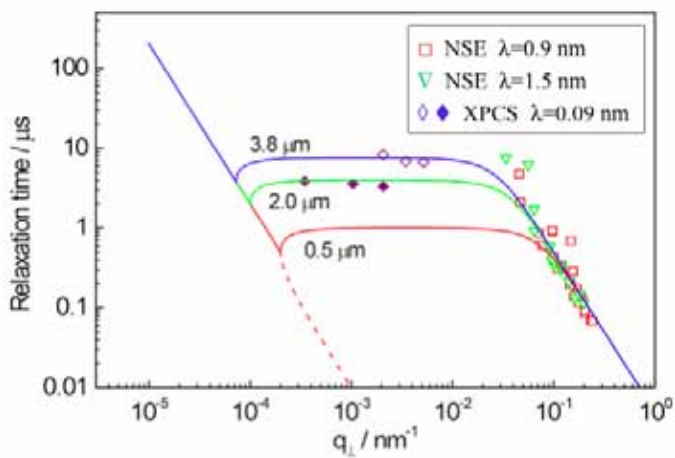

Surely new experiments will follow in the fast expanding field of surfaces and interfaces. May be some words should be said on possible pitfalls concerning NSE experiments. In fact some kind of elastic scatterer is always measured to calibrate the instrument resolution. As we measure directly S(q,t), the deconvolution of the instrumental resolution as is done in omega space, becomes a simple division. Different neutron trajectories explore slightly different field integrals, thus leading to finite instrumental resolution. Special care has to be taken that the elastic scatterer mimics as closely as possible the scattering profile and geometry of the real sample. Indeed the unusually long sample and thin slits introduces very specific correlations between positions and trajectories, which can give artifacts which sometimes very much look like real effects. A closer look on the above shown NSE curves reveals that on the Bragg peak, while very close to 1.0 is not quite within the error bars. After different trials, for the resolution measurement we used a strongly scattering grafoil piece placed exactly where the film was, in the same sample holder with the same slits. The only difference which remained, was that the sharp Bragg peak of the sample actually remonchromatizes the beam better than the 15% original monochromatization. Such effects are particularly dangerous on the side of strong peaks. In our case when the sample was rotated at least 1 degree, the Bragg peak moved off the 2D detector and the scattered intensity was more uniform. + and – rotation gave also identical results, thus we are rather confident that artifacts are at most, but rather smaller than the deviation of the NSE curve from 1.0 on the Bragg peak.

## X-ray Photon Correlation Spectroscopy

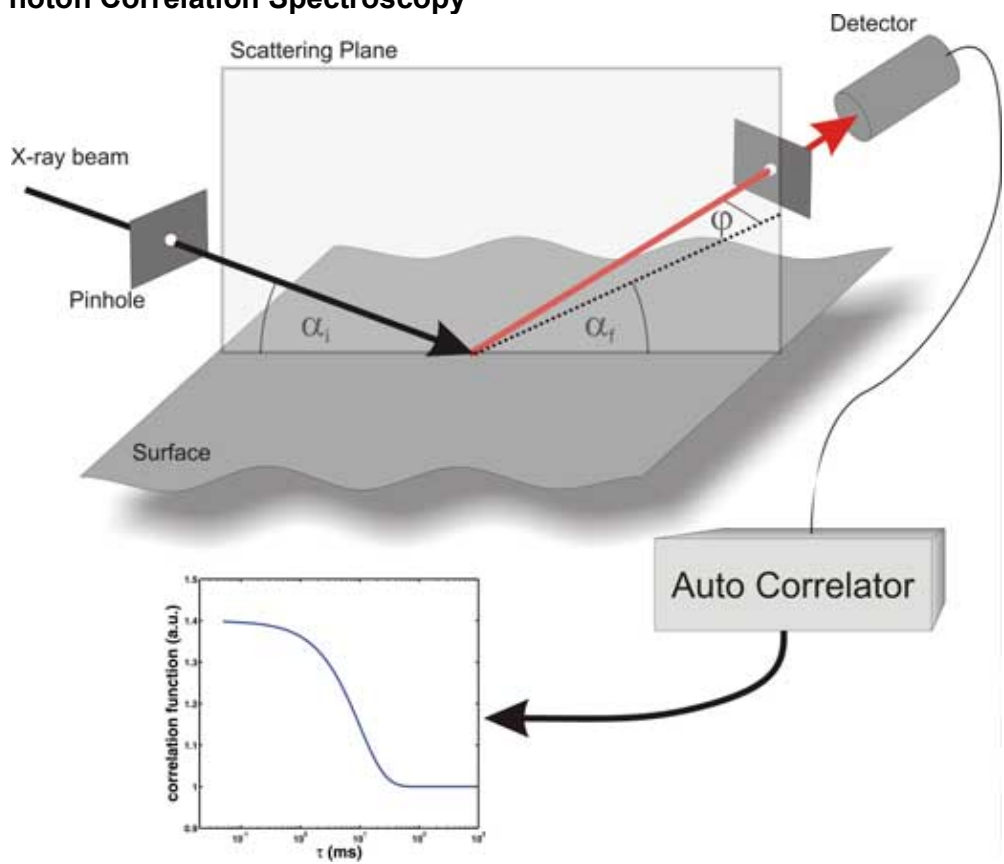

**Figure 9**: Schematic of the XPCS set-up for surface scattering and sketched theoretical curve of a single exponential decay as expected in the intensity autocorrelation from a simple relaxational motion.

Since only a few years ago, it has become possible to study lateral dynamics at surfaces in the most direct and unambiguous way using x-ray photon correlation spectroscopy (8)(9)(10). XPCS requires (partially) coherent x-rays which are only available at the most advanced synchrotron sources. The basic idea of XPCS is to record the intensity autocorrelation of the speckles visible in coherently illuminated matter. This can be done with an excellent wave vector resolution. In measuring the intensity correlation in time, XPCS somewhat resembles to the neutron spin-echo technique in that the intermediate scattering function $S(Q,t)$ rather than $S(Q,\omega)$ is accessed. Surface XPCS takes advantage of the critical angle for total external reflection for x-rays impinging through vacuum or air onto a surface which is greater than zero (see Figure 9). Therefore, it is possible to illuminate surfaces under grazing angles of incidence and detecting a scattering pattern arising from the near-surface region restricted by the 1/e-penetration depth of the evanescent x-ray wave field.

Surface XPCS has the additional advantage that the available coherence volume is projected onto the surface under grazing angles. Therefore, several hundred micrometers can be coherently illuminated along a surface, even though the typical transverse coherence length at 8keV at $3^{rd}$ generation light source is only on the order of 10 $\mu$m. In addition to being surface sensitive, XPCS has the advantage over visible laser light scattering that shorter length scales may be probed and opaque materials may be studied. However, so far the promise of reaching shorter length scales has been severely limited by the photon count rate. Even at the present day most brilliant sources the maximum attainable wave vector transfer is restricted by the limited coherent flux. In addition, there is a principal limit regarding the shortest attainable time scales using XPCS. This limit is caused by the intrinsic time structure of the electron bunches within any synchrotron storage ring and is on the order of a few nanoseconds. Furthermore, because the inelastic or quasi-elastic processes in the sample are recorded through the intensity autocorrelation function, the minimum intensity that is required to obtain meaningful data is essentially proportional to the frequency in the sample to be detected. Even with the advent of free electron x-ray lasers (FELs), a certain time range may remain unexplored, because FELs will emit the radiation in pulses with a pulse frequency much larger than the bunch frequency in synchrotrons, and detectable correlation functions may therefore remain restricted to time scales within a single pulse.

Recently , some of us have performed a first experiment using coherent x-ray radiation at the Troïka 1 beamline of the ESRF to study solid supported lipid membranes (DMPC) as well as charged surfactant bilayers (11)(12). The original idea was to probe collective membrane dynamics on mesoscopic length scales, such as undulation modes. Note that much work in the literature has been dedicated to the anlysis of elastic x-ray scattering in terms of thermal diffuse scattering, without being able to directly evidence the corresponding relaxation time scales. In contrast to the expected smooth scattering curve, averaged out by the fluctuations, we have recorded a pronounced (static) speckle pattern in the diffuse scattering when measuring with coherent x-ray beams. Static speckles were observed at small to medium parallel momentum transfer accessible by scans in the plane of incidence (rocking scan, detector scans), at positions corresponding to the diffuse Bragg sheet where the multilamellar stack dominates the signal. Contrarily,the curves recorded over a range of higher parallel momentum transfer out of the plane of incidence were smooth. This finding might be explained if we assume that in the small Q regime, the diffuse signal stems predominantly from static defects and domain scattering, while at high Q the signal of truly dynamic thermal disorder prevails, which would average out the speckle pattern in time to a continuous curve. However, other effects, such as substrate interactions and geometrical effects may also have to be taken into account. At the same time, we were not able to record an XPCS signal in the high-Q range, possibly because the dynamics was too fast, and the photon count rate too small. This conclusion is supported by numerical estimations based on smectic hydrodynamics, which predicts a strong increase of the frequency at high Q. Contrarily, free standing films also exhibit much slower acoustic modes, which correspond to center of mass movement of the whole film. These slower modes have recently been successfully measured by a combination of XPCS and NSE (13). In smectic films or multilamellar membranes deposited on solid surfaces, the low Q acoustic modes are suppressed by the boundary conditions of the flat substrate.
In summary, a first coherent x-ray experiment has shown that solid supported hydrated and highly fluid lipid bilayer films can yield a static speckle pattern at low-Q. Note that incoherent measurements are not able to distinguish the source and nature of the scatter (static or dynamic), and in the literature to date, the interpretation of the diffuse signal is almost exclusively restricted to thermal fluctuations. The present work points out a caveat for the x-ray line shape analysis of smectic systems from which the membrane elasticity is deduced. Contrarily, inelastic neutron experiments do not suffer from this limitation, since they probe the inelastically scattered neutrons directly, and therefore separate out the elastic scattering. Due to the additional information contained in the energy transfer, the neutrons recorded at a given scattering angle contain much more information than a comparable number of photons. In XPCS the information on dynamics is deduced from photon statistics, making it an inherently "photon hungry" technique. The present example shows that this limitation cannot be overcome only by increasingly brighter light sources. In view of these limits in time scale and competing elastic scattering, it is clearly worth to further develop neutron based instruments and methods to apply inelastic and quasi-elastic neutron scattering to surfaces and interfaces. To this end, collective undulations and bending motions of lipid membranes probed by spin-echo (see previous section) present an encouraging example.

**Conclusion**
In summary we have illustrated the potential relevance of inelastic scattering techniques to study dynamics of surfaces and interfaces, taking the example of planar lipid membranes. The combination of the different techniques, neutron triple-axis, backscattering and spin-echo spectroscopy, as well as x-ray photon correlation spectroscopy maximizes the accessible Q-ω range covering nine decades in energy transfer and spatial dimensions from intermolecular distances to several hundred μm. Dynamics in biomimetic membranes is of particular interest in membrane biophysics to better understand the highly complex dynamics of biological membranes. An understanding of membrane dynamics can also be useful to tailor membrane properties for biotechnology applications.

**Acknowledgements:**
We thank Christoph Ollinger (X-ray physics institute, Göttingen) for help in sample preparation and for an enjoyable collaboration in some of the original studies, Giovanna Fragneto (ILL) for support and collaboration on related studies, Matthias Elender (ILL) for technical and engineering support, and the ILL and the ESRF for allocation of beam time.